# Ferroelectric Nematic Liquid Crystal-Based Silicon Photonic Modulator Demonstrated at 102 Gbit/s PAM-4

Li-Yuan Chiang[(1)], Gianlorenzo Masini[(1)], Rih-You Chen[(2)], Chirag Patel[(1)], Pavel Savechenkov[(1)], Yi-Jen Chiu[(2)], Cory Pecinovsky[(1)], and Jason W. Sickler[(1)]

[(1)] Polaris Electro-Optics, Inc., USA, liyuan.chiang@polariseo.com
[(2)] Department of Photonics, National Sun Yat-sen University, Kaohsiung, Taiwan

**Abstract** *We present the first demonstration of a hybrid integrated silicon and ferroelectric nematic liquid crystal modulator achieving 102 Gbit/s PAM-4 modulation. Operating within the C-band at a 3.5 V DC bias, this Pockels-based Mach-Zehnder modulator exhibits a $V_\pi L(AC)$ of 0.3 V·cm. ©2024 The Author(s)*

**Introduction**

Silicon photonics is increasingly used for a wide range of applications, including optical communications, artificial intelligence, and quantum computing [1]. Optical modulators provide high-speed tunability and are essential components in photonic integrated circuits (PICs) [2]. As the demand for greater data bandwidth continues to grow, the need for optical modulators with higher data rates becomes significant [3]. However, conventional silicon photonic modulators, which typically rely on the plasma dispersion effect, have limited modulation efficiency and bandwidth, posing challenges for realizing next-generation silicon photonic high-speed PICs and systems [1].

Silicon-organic hybrid (SOH) modulators offer a promising solution to advancing modulator efficiency and bandwidth by combining mature CMOS manufacturing with the superior electro-optic (EO) properties of certain organic materials. Among these materials, EO polymers have demonstrated excellent modulation efficiency and bandwidth through the Pockels effect [4,5]. However, issues with scalability, repeatability, and thermal stability remain critical challenges. Nematic liquid crystals (NLCs), though offering strong birefringence for ultra-efficient modulation via molecular reorientation, are limited by the low modulation bandwidth, typically in the kHz range [6,7].

Discovered in 2020 [8], ferroelectric nematic liquid crystals (FNLCs) provide both a high electro-optic coefficient ($r_{33}$) and high bandwidth via the Pockels effect [9,10]. Unlike EO polymers that require poling, FNLCs spontaneously organize into a polar structure. [8]. The linear EO response via the Pockels effect is intrinsic to the FNLC phase. Finally, FNLCs are more scalable and repeatable compared with EO polymers. These characteristics make FNLCs an ideal material for future SOH modulator designs.

This work demonstrates a high-performance hybrid traveling-wave Mach-Zehnder modulator (MZM) integrating FNLC with silicon slot waveguides on a silicon-on-insulator (SOI) platform. The MZM exhibits a modulation efficiency $V_\pi L(AC)$ of 0.3 V·cm at 25 MHz, with a 6-dB bandwidth ($f_{6dB}$) exceeding 67 GHz. Four-level pulse amplitude modulation (PAM-4) transmission of a 102 Gbit/s signal is successfully demonstrated in an FNLC-based device for the first time to our knowledge.

**Design and Fabrication**

Fig. 1(a) illustrates the cross-sectional structure of the hybrid MZM fabricated on a standard SOI wafer. The device features a ground-signal-ground-signal-ground (GSGSG) traveling-wave electrode, asymmetrically loaded with two doped-silicon slot waveguide phase shifters. FNLC serves as the dielectric cladding material [11] as well as the actuated material in the slot of the slot waveguide [9]. Due to the high spontaneous polarization of FNLC, the molecules orient with the direction of the applied DC bias [8]. High-speed modulation via the Pockels effect of the FNLC in the slot regions is achieved through electrical bias in the horizontal directions and operating the device using TE0 optical mode.

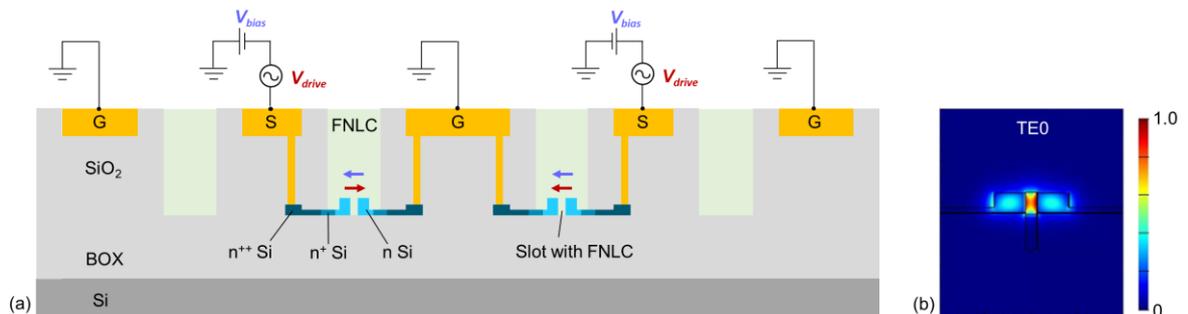

**Fig. 1:** (a) Cross-sectional schematic of the MZM. (b) Optical mode field distribution of a Si-FNLC slot waveguide.

To support push-pull driving, the two signal traces are applied with either opposite $V_{bias}$ and an in-phase modulating signal ($V_{drive}$) or the same $V_{bias}$ with out-of-phase $V_{drive}$. Fig. 1(a) illustrates the former case. Fig. 1(b) shows a simulated TE0 mode field distribution in the Si-FNLC slot waveguide. Efficient electro-optic modulation is facilitated by the concentrated optical and electrical fields in the slot region [12].

The device was fabricated in a commercial foundry using a standard active silicon photonics MPW including cladding oxide open process. No process optimization was conducted for this work. FNLC loading was performed in-house through an integration process post-foundry.

Cross-sectional transmission electron microscopy (TEM) analysis was conducted before FNLC loading to measure the dimensions of the fabricated waveguide structures. The measured slot width is approximately 125 nm and the phase shifters are 1 mm long. There is a 350 nm oxide overetch into the buried oxide (BOX) layer at the slot regions to ensure slot clearance from the oxide open process.

**Experimental Results**

The FNLC material used in this work was PM616, a proprietary mixture designed for large second-order nonlinear optical susceptibility. The modulator was operated in the C-band. AC characterization with a 25 MHz sinusoidal driving signal was used to measure the half-wave voltage ($V_\pi$) and preclude low-frequency molecular responses. Fig. 3(a) shows measured data of the input signal (dotted curve) at 5 Vpp and the corresponding modulated waveform (solid curve). A $V_\pi(AC)$ of 3 V was extracted by fitting modulated waveforms at different wavelengths. The results correspond to a modulation efficiency of $V_\pi L(AC)$ = 0.3 V·cm. Using the $V_\pi$ results, the waveguide dimensions from TEM analysis, and Lumerical MODE simulations, an $r_{33}(in\text{-}device)$ of 29.5 pm/V was extracted.

A $V_{bias}$ of 3.5 V was maintained during both AC & RF characterization to ensure proper orientation of the FNLC domain within the slot regions. The $V_{bias}$ magnitude was selected to exceed the FNLC's alignment saturation threshold for the specific slot width. Owing to the thermodynamic stability of the FNLC phase and its high spontaneous polarization, the FNLC domain reorientation does not require a high electric field approaching dielectric breakdown, as is necessary for the poling process of EO polymer modulators [13]. As a result, FNLC offers a scalable and repeatable solution for silicon-organic hybrid modulators, surpassing the limitations of EO polymers. Pure passive alignment through physical constraints is a potential alternative for controlling FNLC directions [14, 15].

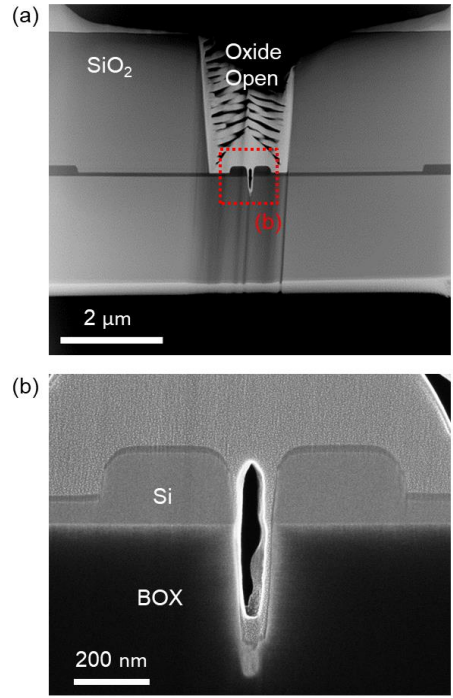

**Fig. 2:** (a)-(b) Cross-sectional TEM images of the device before FNLC loading.

Fig. 3(b) displays a measured transmission spectrum of the MZM, showing an FSR of 7.5 nm and an extinction ratio of 21.8 dB. The Si-FNLC phase shifters exhibited a propagation loss of approximately -6.5 dB/mm. Insertion loss can be reduced in future iterations through process optimization, advanced techniques for reducing sidewall roughness [16,17], and using shorter phase shifter designs integrated with next-generation FNLC mixtures with higher $r_{33}$ values.

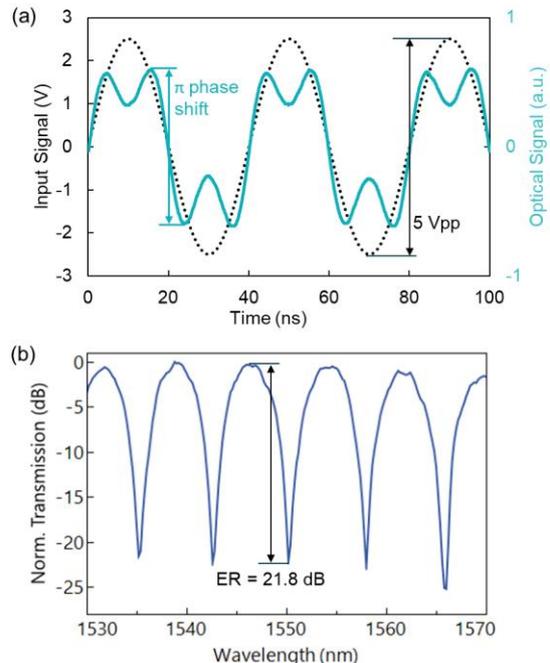

**Fig. 3:** Measured (a) AC modulation at 25 MHz and (b) optical transmission spectrum.

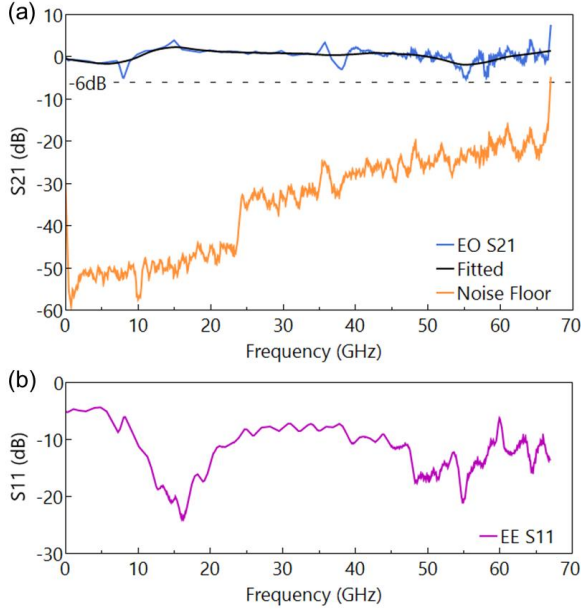

**Fig. 4:** Measured (a) small-signal EO response ($S_{21}$) and (b) electrical reflection ($S_{11}$).

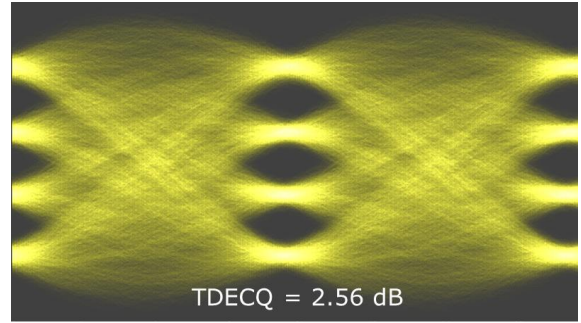

**Fig. 5:** Observed eye diagram of 102 Gbit/s (51-GBd) PAM-4 signal equalized with 42-tap linear equalizer.

The modulator's S-parameters were characterized using a 67 GHz PNA Network Analyzer (Agilent/Keysight E8361C), with an RF splitter and in-phase small signals applied simultaneously to both signal traces of the GSGSG traveling-wave electrode. The results are plotted in Fig. 4. No roll-off was observed in the EO $S_{21}$ measurements, as shown by the fitted curve in Fig. 4(a). The measured $f_{6dB}$ exceeds 67 GHz and is limited by instrumentation. The noise floor was measured immediately after obtaining the EO $S_{21}$ by turning off the laser while maintaining the same testing setup. Although the noise in the $S_{21}$ measurement became significant at frequencies above 45 GHz due to RF losses from the cabling and components, the $S_{21}$ signals remained approximately 20 dB above the noise floor up to 65 GHz. While some unwanted ripple was observed in the $S_{21}$ data, and the $S_{11}$ results indicated higher reflections at lower frequencies, these issues can be mitigated in future designs by optimizing the transmission line design.

Fig. 5 presents the observed eye diagram of the modulated light when the MZM was driven by a 102 Gbit/s (51-GBd) PAM-4 signal with a pseudo-random binary sequence (PRBS) equal to 11. There are eight waveforms overlapped in the eye diagram. Push-pull driving was implemented with a 1.9 Vpp differential input signal (before equalization) and $V_{bias}$ of 3.5 V. The open eyes were achieved using a 7-tap feed-forward equalizer (FFE) on the transmitter side and 42-tap FFE on the receiver side. A transmitter dispersion eye closure quaternary (TDECQ) of 2.56 dB was measured.

The high-speed testing results confirm Pockels-based modulation using an FNLC-integrated silicon slot waveguide structure. The efficient modulation enabled by the Pockels effect positions FNLC as a promising candidate for achieving high data rates beyond 102 Gbit/s, with a low $V_\pi$, compact footprint, scalability, and repeatability.

## Conclusions

We have successfully demonstrated the first hybrid Si-FNLC modulator capable of 102 Gbit/s PAM-4 modulation, achieving a $V_\pi L(AC)$ of 0.3 V·cm and $f_{6dB}$ >67 GHz. The integration of FNLC into a silicon slot waveguide structure enables efficient and high-speed modulation through the Pockels effect. The device was fabricated using a standard silicon photonics MPW run. A monodomain of FNLC molecules over the length of the device can be established and maintained with modest DC fields, due to coupling with the large spontaneous polarization intrinsic to the ferroelectric nematic phase. Avoiding the large poling fields required to induce polar order enables an easily manufacturable integration process as well as a thermodynamically stable active state. These advantages position Si-FNLC modulators as a promising solution to surpass the performance limitations of conventional plasma dispersion-based silicon photonic modulators.

## Acknowledgments
The eye diagram results in this work were made possible with the support of Multilane. Special thanks to Jana El Husseini and Chafik Hoayek for their discussion and technical support.